\documentclass[12pt]{iopart}

\usepackage{float}
\usepackage{booktabs}
\usepackage{graphicx}
\usepackage{dcolumn}
\usepackage{bm}
\usepackage{cite}

\begin{document}

\title{Comparing discriminating abilities of evaluation metrics in link prediction}

\author{Xinshan Jiao$^1$, Shuyan Wan$^1$, Qian Liu$^2$, Yilin Bi$^1$, Yan-Li Lee$^3$, En Xu$^4$, Dong Hao$^1$, Tao Zhou$^1$}

\address{$^1$ CompleX Lab, School of Computer Science and Engineering, University of Electronic Science and Technology of China, Chengdu 610054, China}
\address{$^2$ School of Economics and Management, Harbin Institute of Technology (Shenzhen), Shenzhen 518055, China}
\address{$^3$ School of Computer and Software Engineering, Xihua University, Chengdu 610039, China}
\address{$^4$ Department of Physics, Hong Kong Baptist University, Kowloon 999077, Hong Kong SAR}

\ead{zhutou@ustc.edu}
\vspace{10pt}
\begin{indented}
\item[]January 2024
\end{indented}

\begin{abstract}
Link prediction aims to predict the potential existence of links between two unconnected nodes within a network based on the known topological characteristics. Evaluation metrics are used to assess the effectiveness of algorithms in link prediction. The discriminating ability of these evaluation metrics is vitally important for accurately evaluating link prediction algorithms. In this study, we propose an artificial network model, based on which one can adjust a single parameter to monotonically and continuously turn the prediction accuracy of the specifically designed link prediction algorithm. Building upon this foundation, we show a framework to depict the effectiveness of evaluating metrics by focusing on their discriminating ability. Specifically, a quantitative comparison in the abilities of correctly discerning varying prediction accuracies was conducted encompassing nine evaluation metrics: Precision, Recall, F1-Measure, Matthews Correlation Coefficient (MCC), Balanced Precision (BP), the Area Under the receiver operating characteristic Curve (AUC), the Area Under the Precision–Recall curve (AUPR), Normalized Discounted Cumulative Gain (NDCG), and the Area Under the magnified ROC (AUC-mROC). The results indicate that the discriminating abilities of the three metrics, AUC, AUPR, and NDCG, are significantly higher than those of other metrics. 
\end{abstract}

\vspace{2pc}
\section{Introduction}
Link prediction represents a highly vibrant research direction within the realm of network science\cite{survey}. Over the past decade, numerous pioneering works in link prediction have emerged\cite{survey, Kleinberg, Clauset, Guimera, Leskovec}, and  it has also been found that link prediction can be applied in various fields, such as life sciences, information security, social network analysis, and transportation planning\cite{Almansoori, Huang, Wu, Aiello, Wang}. In social media platforms, the application of link prediction facilitates a more accurate comprehension and forecasting of  relationship developments within social networks\cite{Adamic}. The link prediction techniques enhance the precision and personalization level of recommender systems\cite{Medo}, thereby increasing user stickiness\cite{Kleinberg}. In the domain of life sciences, a significant proportion of interactions remain unobserved. For instance, in protein-protein interaction networks, approximately only 20\% of interactions among yeast proteins are known, and engaging in blind experimental validation of potential interactions will lead to substantial resource wastage. Conversely, employing link prediction techniques enables the probable interactions for subsequent validation, thereby significantly enhancing efficiency and reducing experimental costs\cite{Yu, Sulaimany, Lei, Barabasi}. 

As of the present time, the majority of research endeavors pertaining to link prediction have been primarily centered on designing novel algorithms or refining existing algorithms. Given the inherent randomness and structural complexity of real-world networks, the choice of efficacious algorithms highly depends on the specific network under consideration, thus presenting both challenges and opportunities in algorithmic research\cite{survey, Zhou, Zhou2}. The careful selection of pertinent evaluation metrics for the accurate assessment of algorithm performance stands as an indispensable cornerstone within the domain of link prediction. However, the previous study on this pivotal issue has been somewhat arbitrary, with many researchers predominantly resorting to classical evaluation metrics such as AUC, BP, and AUPR. Recently, several scientists have conducted critical reevaluations of this foundational issue, with a particular focus on commonly employed metrics such as Precision and AUC\cite{Powers, Saito, Hand}. For instance, Yang, Lichtenwalter, and Chawla\cite{yy} argued that, when evaluating link prediction performance, the precision-recall curve might provide better accuracy than AUC. Similarly, Austin\cite{Austin} emphasized the need to reevaluate the sole reliance on AUC as a benchmark for model efficacy, especially in the prediction of species distribution. Additionally, Lobo {\it et al}\cite{lobo} also questioned AUC and offered reasons against its use. In the pursuit of how to evaluate algorithms for imbalanced classification and a nuanced characterization of the differences between algorithms, some researchers introduced innovative evaluation metrics, such as AUC-mROC\cite{Muscoloni}.

In this study, we introduce an artificial network model paired with a corresponding link prediction algorithm. Within this algorithm, we incorporate a parameter to adjust the noise intensity, wherein increased noise reduces the algorithm's prediction accuracy. Consequently, we can use a single parameter to monotonically and continuously adjust the prediction accuracy of the algorithm. If a metric can depict variances in prediction accuracy across diverse noise intensities accurately, it shows that the metric has strong discriminating ability and can distinguish the pros and cons of different algorithms. We employ the recently proposed methodology for quantifying metric performance based on discriminating ability\cite{Discriminating}, undertaking a quantitative comparative analysis encompassing nine evaluation metrics pertinent to link prediction algorithms. This work facilitates to elucidate current controversy and confusion within this domain, offering guide to design novel evaluation metrics. Furthermore, the findings derived from this work hold broader implications and provide valuable references for addressing more generalized classification challenges. 

\section{Problem Description}

Let $G(V,E)$ denote a network, where \(V\) represents the set of nodes and \(E\) represents the set of links. A link is drawn between two nodes if there exists a certain relationship or interaction between them \cite{structure}. For instance, in social networks, users can be represented as nodes, and friendships as connecting links. This study considers the simplest type of networks, ignoring the weight and directionality of links, and disallowing multiple links and self-connected links. Denoting the size of \(G\) as the cardinality of its node set, say \(N = |V|\), and the set of all potential links within \(G\) as \(U\). Evidently, \(|U| = N(N-1)/2\). There may be some existing links in U-E but not yet being observed. For instance, in biological networks, numerous interactions remain undiscovered, commonly referred to as missing links. Alternatively, with the evolution of networks, new links may appear, often termed as future links. The primary objective of link prediction algorithms is to predict, based on the observed link set \(E\), which among the potential links in \(U-E\) are likely to be missing links or future links. 

To facilitate the training of models and validation of algorithms, we need to partition the observed link set \(E\) into training set \(E^T\) and testing set \(E^P\), ensuring that \(E = E^T \cup E^P\) and \(E^T \cap E^P = \emptyset\). The links in the training set are considered known, whereas those in the testing set are regarded as unknown and serve as the basis for algorithmic validation. Evidently, an efficacious algorithm should identify links in \(E^P\) as having higher likelihoods of being either missing links or future links among all presumed unknown links \(U-E^T\). In practice, when forecasting missing links, a subset \(E^P\) is typically constituted by randomly selecting links from \(E\), and when predicting future links, a subset \(E^P\) is often constituted by choosing links appearing later. This study predominantly focuses on the former scenario. It is noteworthy that link prediction is also a typical binary classification problem, so, most of the evaluation metrics tailored for binary classification  problems can be seamlessly adapted to evaluate the efficacy of link prediction algorithms. 

\section{Evaluation metrics}

Evaluation metrics can be broadly categorized into threshold-dependent metrics and threshold-free metrics. Threshold-dependent metrics produce results that are contingent on the chosen threshold parameters, whereas threshold-free metrics are independent of any threshold parameters. Given that the selection of thresholds often entails an ad hoc approach, which may not be universally applicable, threshold-dependent evaluation metrics are frequently perceived as lacking in persuasiveness for addressing generalized problems. In contrast, threshold-free metrics are generally favored, unless the selection of thresholds is tied to the specific problem rather than being arbitrarily designated by researchers. This work considers nine evaluation metrics, including four threshold-dependent metrics and five threshold-free metrics.

\subsection{Threshold-Dependent Metrics}
Common threshold-dependent metrics include Precision@k\cite{precision}, Recall@k\cite{yy}, F1-Measure\cite{f1}, and MCC\cite{Matthews}. Before introducing these specific metrics, we first review the confusion matrix in binary classification problems. Within the confusion matrix, all samples are classified into four categories based on whether they are positive samples (corresponding to missing links \(E^P\) in link prediction) or negative samples (corresponding to  non-existent links \(U-E\) in link prediction), and whether they are correctly predicted. These four categories are: true positive (TP), where a positive sample is correctly predicted as positive; false positive (FP), where a negative sample is incorrectly predicted as positive; true negative (TN), where a negative sample is correctly predicted as negative; and false negative (FN), where a positive sample is incorrectly predicted as negative. 

Precision is delineated as the ratio of samples accurately predicted as positive by the algorithm that are indeed positive samples. Without loss of generality, a link prediction algorithm can rank all links in the set \(U-E^T\) in descending order based on their likelihoods of being missing links. The algorithm may then consider the top-\(k\) links as missing links (positive samples), while the remaining links are considered as non-existent links (negative samples). In this context, the parameter \(k\) serves as a typical threshold parameter. Selecting the top-\(k\) links as predicted missing links is equivalent to setting a likelihood threshold and considering links with likelihood scores higher than this threshold as predicted missing links. Once \(k\) is determined, precision is calculated as the proportion of potential links ranked within the top \(k\) that are indeed missing links, as
\begin{equation}
    Precision@k=\frac{TP@k}{TP@k+FP@k}=\frac{TP@k}{k}, 
    \label{Precision@k}
\end{equation}
where \( TP@k \) and \( FP@k \) respectively refer to the number of missing links and non-existent links among the top-\( k \) potential links as ranked by the algorithm. 

Recall is defined as the proportion of positive samples that are correctly predicted as positive by the algorithm. Clearly, given a specific algorithm, as \( k \) increases, \( Recall@k \) is monotonically non-decreasing. When \( k = |U-E^T| \), i. e., the algorithm predicts all potential links as missing links, \( Recall@k = 1 \). The formula to compute \( Recall@k \) is as follows:
\begin{equation}
  Recall@k=\frac{TP@k}{TP@k+FN@k}=\frac{TP@k}{|E^P|}. 
  \label{Recall@k}
\end{equation}

As the parameter \( k \) varies, Precision and Recall generally exhibit inverse trends. To balance the influence of both metrics and provide a more holistic evaluation of algorithm performance, the F1-Measure computes the harmonic mean of Precision and Recall, as
\begin{equation}
    F1=\left( \frac{\frac{1}{Precision}+\frac{1}{Recall}}{2} \right)^{-1}=\frac{2 \cdot Precision \cdot Recall}{Precision+Recall}. 
\end{equation}

For the sake of clarity, we omit $@k$ in equation (3) and some later equations, given that no ambiguity will be introduced. The values of Precision, Recall and F1-Measure are all in the interval [0, 1]. 

MCC is utilized to depict the correlation between actual outcomes and predicted results, taking into account the values of TP, FP, TN, and FN. Due to its balanced nature, even when there is a significant disparity in the sample sizes between the two classes, MCC can effectively reflect the algorithm's performance. The formula for MCC is as follows:

\begin{equation}
    MCC=\frac{TP\cdot TN-FP \cdot FN }{\sqrt{(TP + FP)\cdot (TP + FN) \cdot (TN + FP)\cdot (TN + FN)} }. 
    \label{MCC}
\end{equation}
The range of MCC is $[-1,1]$, where $MCC=1$ indicates prefect prediction (corresponding to \( FP = FN = 0 \)). Conversely, \( MCC = -1 \) signifies entirely erroneous predictions. Random classification is corresponding to \( MCC \approx 0 \). According to equation (\ref{MCC}), TP, FP, TN, and FN carry equal importance, suggesting that even if the positive and negative samples are interchanged, the value of MCC remains unchanged, underscoring its symmetric nature. 

\subsection{Threshold-Free Metrics}

Threshold-free metrics never require any effort on determining appropriate thresholds, and they also circumvent the issue that different thresholds lead to different winners. Well-know threshold-free metrics include BP\cite{precision}, AUC\cite{auc}, AUPR\cite{aupr}, and NDCG\cite{ndcg}. This work will also analyze a recently proposed metric called AUC-mROC\cite{Muscoloni}. 

BP represents the intersection of the Precision@k and Recall@k curves, specifically when the threshold \( k \) equals the size of the testing set (i. e., \( k = |E^{P}| \)), as
\begin{equation}
  \label{Precision}
  BP=\frac{TP@|E^{P}|}{\left |E^{P} \right |}. 
\end{equation}

\begin{figure}[H]
  \centering
  \includegraphics[width=1\textwidth]{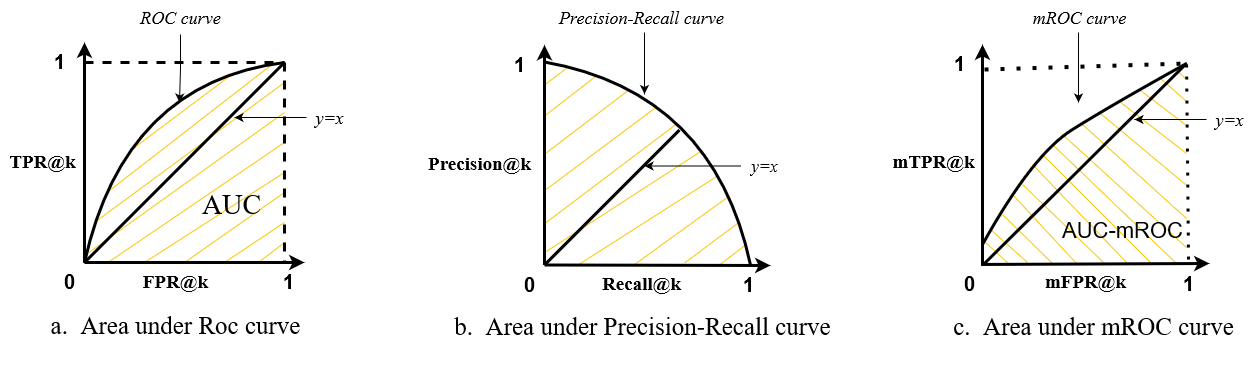}
  \caption{Illustration about (a) AUC, (b) AUPR and (c) AUC-mROC}
  \label{ROC}
\end{figure}

AUC represents the area under the receiver operating characteristic (ROC) curve. For each threshold \( k \), there exists a corresponding point on the ROC curve. The x-coordinate of this point is the false positive rate at \( k \), denoted as \( FP@k/k \), and the y-coordinate is the true positive rate at \( k \), denoted as \( \frac{TP@k}{k} \). By varying the threshold \( k \) from small to large values, the ROC curve can be obtained. For the specific task of link prediction, there is a more straightforward method to plot the ROC curve. We firstly sort all \( |U-E^T| \) potential links based on their predicted likelihoods in descending order, then we start the origin \( (0, 0) \) and sequentially scan these \( |U-E^T| \) potential links. If a missing link is encountered, we move upwards by \( \frac{1}{|E^P|} \), while if a non-existent link is encountered, we move right by \( \frac{1}{|U-E|} \). Upon completing this scan, the ROC curve is obtained that spans from \( (0, 0) \) to \( (1, 1) \). If the likelihoods of potential links are entirely randomly assigned, the ROC curve would approximate the diagonal, with an AUC close to 0.5. Generally, an AUC value provided by an algorithm will range between 0.5 and 1; a value closer to 1 indicates better prediction performance. 

The value of AUC can be intuitively interpreted as the probability that a randomly chosen positive sample (missing link) has a higher predicted likelihood than a randomly chosen negative sample (non-existent link). This intuitive interpretation is a distinct advantage of the AUC metric. We assume that in the sorted sequence of \( |U-E^T| \) potential links, the positions of the \( |E^{P}| \) missing links are \( r_{1} < r_{2} < r_{3} < \dots < r_{|E^{P}|} \). Then, prior to the $i$th missing link, there are \( r_{i} - i \) non-existent links. Therefore, when comparing the $i$th missing link to all \( |U-E| \) non-existent links, it will lose against \( r_{i} - i \) of them. In other words, its winning probability is \( 1 - (r_{i} - i) / |U-E| \). Accordingly, the AUC can be calculated by averaging these winning probabilities across all missing links, as
\begin{equation}
    AUC=\frac{1}{\left | E^{P} \right | }\sum_{i=1}^{\left | E^P \right | } \left ( 1-\frac{r_{i}-i}{\left | U-E \right | } \right ) =1-\frac{\left \langle r \right \rangle }{\left | U-E \right | }+\frac{\left | E^{P} +1\right | }{2\left | U-E \right | },
\end{equation}
where \( \left \langle r \right \rangle = \sum_{i=1}^{\left | E^P \right | } r_{i} / |E^{P}| \). Given that most real networks are sparse\cite{sparsenets}, i. e., \( |E^P| / |U-E| \ll 1 \), it can be approximated as:

\begin{equation}
  AUC \approx 1-\frac{\left \langle r \right \rangle }{\left | U-E^{T} \right | },
  \label{AUC}
\end{equation}
which is also known as ranking score in some previous literature\cite{rankingscore} AUPR represents the area under the Precision-Recall curve. The PR curve is constructed by plotting Precision (on the y-axis) against Recall (on the x-axis) for various threshold values. For a given threshold \( k \), the corresponding point on the PR curve is \( \left( \frac{TP@k}{|E^{P}|}, \frac{TP@k}{k} \right) \). When \( k \) takes its maximum value \( |U-E^{T}| \), the PR curve ends at the point \( \left( 1, \frac{|E^{P}|}{|U-E^{T}|} \right) \). Similar to equation (6), AUPR can be expressed as:
\begin{equation}
  AUPR=\frac{1}{2\left | E^{P} \right | }\left ( \sum_{i=1}^{\left | E^{P} \right | } \frac{i}{r_{i}} +\sum_{i=1}^{\left | E^{P} \right | } \frac{i}{r_{i+1}-1}\right ), 
  \label{AUPR}
\end{equation}
where \( r_{|E^{P}|+1} \) is defined as \( |U-E^{T}| + 1 \). 

Discounted cumulative gain (DCG) considers that the importance of positions in the ranking of potential links is not uniform. If an algorithm ranks missing links higher up in the list, it receives a higher score. Conversely, if these missing links are ranked lower, their scores are discounted. Specifically, DCG employs a logarithmic discounting mechanism, as

\begin{equation}
  DCG=\sum_{i=1}^{\left | E^P  \right | } \frac{1}{\log_{2}{(1+r_i)}}. 
  \label{DCG}
\end{equation}
Note that equation (9) and equation (7) are quite similar. However, in the approximate definition of AUC, the contribution of a missing link ranked at position \( r \) is \( 1 - \frac{r}{\left | U-E^{T} \right | } \), while in the definition of DCG, the contribution of a missing link at position \( r \) is \( \frac{1}{\log_{2}(1+r)} \). Clearly, as \( r \) increases, the corresponding contribution in DCG diminishes more rapidly. For instance, if there are a total of 10, 000 samples to be predicted, a positive sample at the top rank contributes a score of 1 to both AUC and DCG. However, a positive sample ranked at \( r = 5000 \) contributes a score of approximately 0.5 to AUC but only about 0.08 to DCG. While DCG can be used to compare algorithm performances, its absolute value lacks meaning. To address this challenge and enable cross-dataset comparisons, normalization can be implemented by dividing by the maximum possible value of DCG. Clearly, when all missing links are precisely ranked in the top \( |E^P| \) positions, DCG attains its maximum possible value. Consequently, the corresponding NDCG is given by
\begin{equation}
  NDCG=\sum_{i=1}^{\left | E^P  \right | } \frac{1}{\log_{2}{(1+r_i)}} 
\bigg/ \sum_{r=1}^{\left | E^P  \right | } \frac{1}{\log_{2}{(1+r)}}. 
  \label{NDCG}
\end{equation}

The AUC-mROC\cite{Muscoloni} applies the idea of NDCG to optimize AUC. Specifically, it transforms both axes of the ROC curve using logarithmic transformations. After the transformation, the horizontal and vertical coordinates are defined as \( mFPR@k = \log_{(1+|U-E|)}{(1+FP@k)} \) and \( mTPR@k = \log_{(1+|E^{P}|)}{(1+TP@k)} \) respectively. The AUC-mROC represents the area under this transformed curve. 

\section{Discriminating Ability}
In this study, a simple method is adopted to generate an artificial network \( G(V, E) \)\cite{Garcia}. For any node pair \( (i, j) \) in \( G \) $(i$, $j\in V$, and $1\le i< j\le N)$, the likelihood to form a link is denoted as \( q_{ij} \). These likelihood values are independently generated from a uniform distribution \( U(0, q_{max}) \), where the parameter \( q_{max} \) can be utilized to control the linking density. Once all the likelihood values are generated, links between node pairs are established or not based on their corresponding likelihoods. For instance, if the likelihood value for a link is \( q \), then the probability of establishing this link is \( q \), and the probability of not establishing the link is \( 1-q \). In this model, if all likelihoods are known, the optimal prediction algorithm would set the likelihood for \( (i, j) \) as \( s_{ij} = q_{ij} \) \cite{Garcia}. 

\begin{figure}[H]
  \centering
  \includegraphics[width=0.9\textwidth]{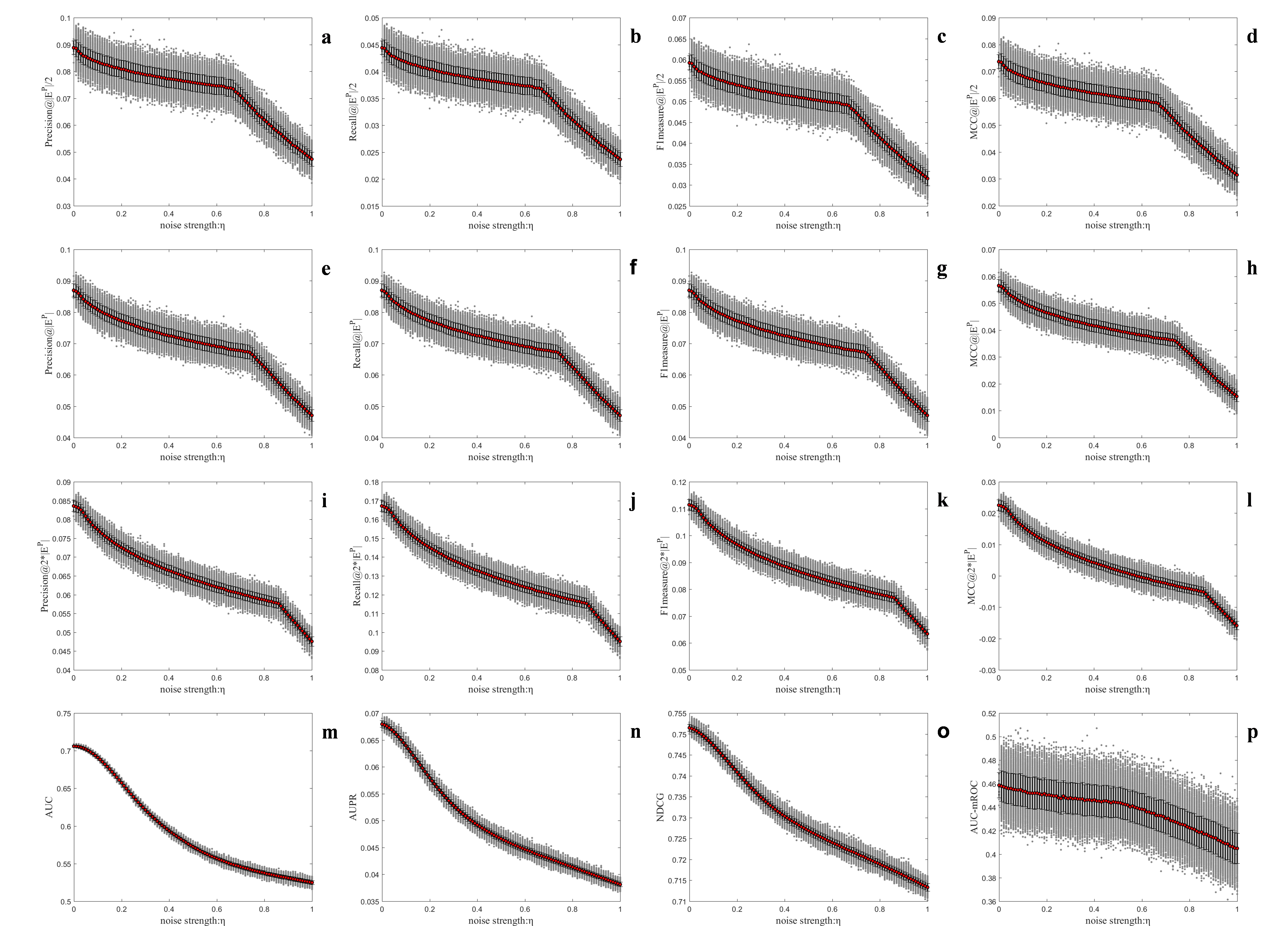}
  \caption{How the value of a metric varies with changing noise. (a)-(d), (e)-(h) and (i)-(l) respectively represent the results when the thresholds for Precision, Recall, F1-Measure, and MCC are set to \( |E^P|/2 \), \( |E^P| \), and \(2|E^P|\). (m)-(p) depict the results for AUC, AUPR, NDCG, and AUC-mROC. The gray points represent the simulation of given values in single runs, the red points represent the average values of given noise intensities, and the error bars indicate the corresponding standard deviations. }
    \label{error}
\end{figure}

\begin{figure}[H]
  \centering
  \includegraphics[width=0.9\textwidth]{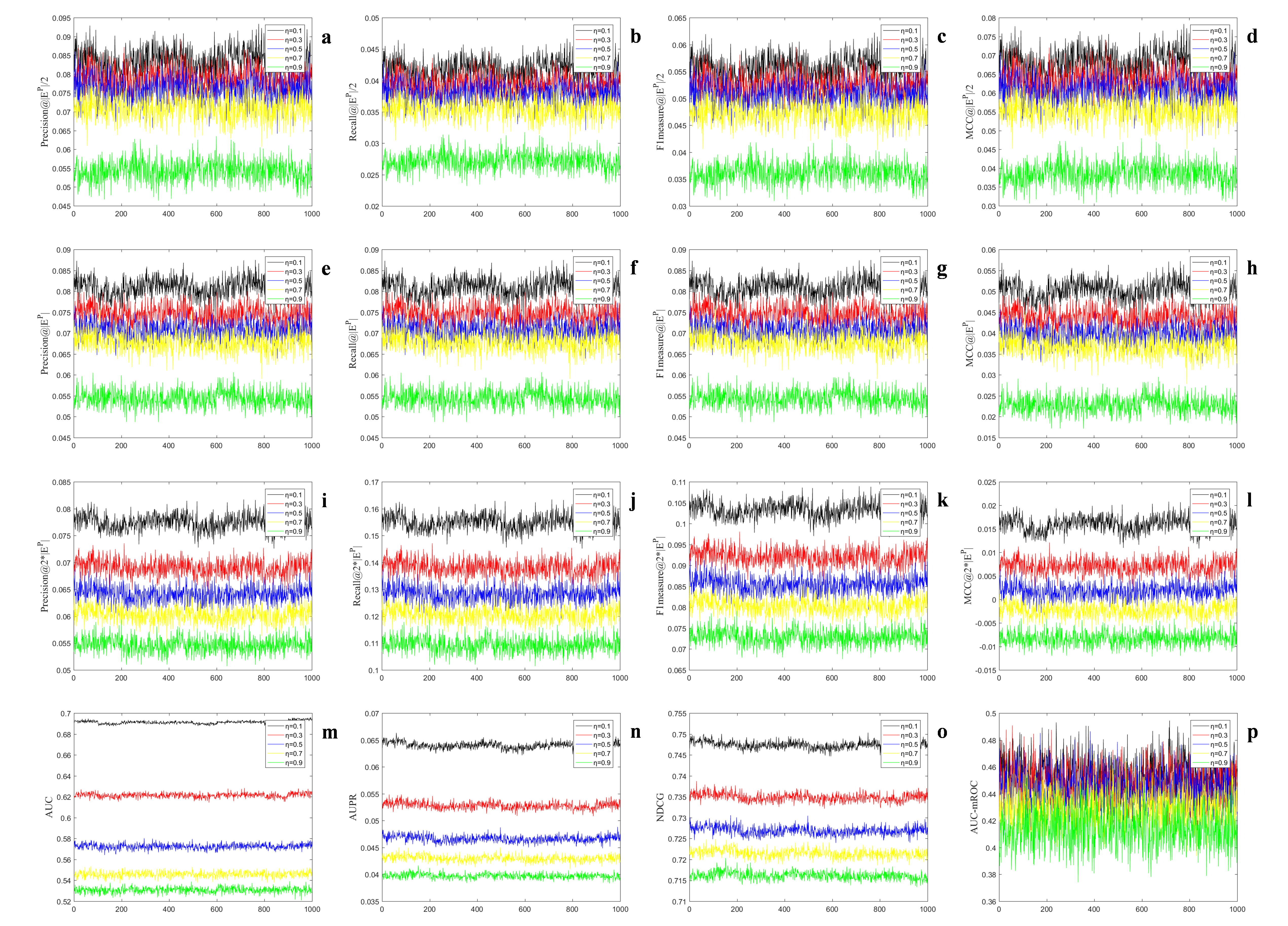}
  \caption{The values of the evaluation metrics under different noise intensities, where $\eta=0.1$, $\eta=0.3$, $\eta=0.5$, $\eta=0.7$ and $\eta=0.9$. The x-axis represents the number of runs, and the y-axis represents the values of the evaluation metrics in the 1000 runs. (a)-(d), (e)-(h) and (i)-(l) represent the results when the thresholds for Precision, Recall, F1-Measure, and MCC are \( |E^P|/2 \), \( |E^P| \), and \( 2|E^P|\), respectively. (m)-(p) denote the results for AUC, AUPR, NDCG, and AUC-mROC. }
  \label{metric}
\end{figure}

By introducing a parameter that depict the noise, we can continuously and monotonically adjust the prediction accuracy of the algorithm. Let \( \Omega(\eta) \) represents an algorithm with noise \( \eta \). The likelihood value \( s_{ij} \) provided by this algorithm for any node pair \( (i, j) \) in \( U-E^T \) is \( s_{ij} = p_{ij} + n_{ij} \), where the noise term \( n_{ij} \) is sampled from a uniform distribution \( U(-\eta, \eta) \). Clearly, as \( \eta \) increases, the prediction accuracy of the algorithm decreases. Given an evaluation metric \( M \) that operates on the algorithm, if there are two noise parameters \( \eta_1 \) and \( \eta_2 \) with \( \eta_1 < \eta_2 \), then \( M(\Omega(\eta_{1})) > M(\Omega(\eta_{2})) \) should hold. If in an experiment, the metric \( M \) indeed satisfies \( M(\Omega(\eta_{1})) > M(\Omega(\eta_{2})) \), we say it correctly distinguishes the performance difference of the algorithm; otherwise, \( M \) fails to do so. Clearly, the smaller \( \eta_2 - \eta_1 \) is, the greater the probability of incorrection. Suppose \( \eta_{1} < \eta_{2} \), and in \( X \) independent comparisons, there are \( x \) comparisons with the result \( M(\Omega(\eta_{1})) \leq M(\Omega(\eta_{2})) \). Then, the p-value for the noise parameters \( (\eta_1, \eta_2) \) given \( M \) is defined as \( p(\eta_{1}, \eta_{2}) = x/X \). When \( p(\eta_{1}, \eta_{2}) \) is less than a pre-defined significance level \( p^* \), we conclude that the metric \( M \) is capable of distinguishing between the algorithms \( M(\Omega(\eta_{1})) \) and \( M(\Omega(\eta_{2})) \). Setting \( p(\eta_{1}, \eta_{2}) = p(\eta_{2}, \eta_{1}) \), if \( \eta_1 > \eta_2 \), and \( p(\eta_{1}, \eta_{2}) = 0.5 \), if \( \eta_1 = \eta_2 \), then \( p(\eta_{1}, \eta_{2}) \) is symmetric. This matrix is referred to as the discrimination matrix, denoted as \( P = p(\eta_{i}, \eta_{j}) \). By contrasting the \( P \)-matrix corresponding to different evaluation metrics, we can intuitively assess the discrimination ability of different evaluation metrics \cite{Discriminating}. 

\section{Results}

In the simulation, we set the number of nodes in the network to \(N=1000\), the parameter of the uniform distribution \(U(0, q_{max})\) is \(q_{max}=0.5\), the proportion of the prediction set is \(|E^{P}|:|E^{T}|=1:9\). For threshold-dependent metrics, we assume that the predicted links are the top-\(k\) links ranked by the value \(s_{ij}\) and mainly demonstrate the cases of \(k=|E^P|/2\), \(k=|E^P|\), and \(k=2|E^P|\). For each set of parameters, we randomly generate 10 networks, and indenpendently run 100 simulations for each network. Due to the simplicity of the model and the independence between the likelihoods, as long as \(N^2\) is sufficiently large and \(q_{max}\) is not too small or too large, the results are similar. Figure \ref{error} shows the situation where the values of different metrics change with the increase of noise intensity. It can be seen that the values of all metrics decrease overall as the noise increases, but some metrics have large fluctuations. Intuitively, metrics with large fluctuations are difficult to distinguish the performance differences of algorithms when the difference is small. In figure \ref{error}, it can be seen that the overall fluctuation of a threshold-dependent metric is generally greater than that of AUC, AUPR, or NDCG. Among all metrics, AUC-mROC has the largest fluctuation. 

Figure \ref{metric} presents the outcomes from 1000 runs at noise intensities of \(\eta=0. 1\), \(\eta=0. 3\), \(\eta=0.5\), \(\eta=0. 7\), and \(\eta=0.9\). Clearly, if there are noticeable gaps between the curves of different noise intensities for a specific metric, it indicates that the metric can differentiate the performance differences of algorithms under various noise intensities. In figure \ref{metric}, it is evident that AUC, AUPR, and NDCG can effectively distinguish between adjacent noise intensities (i. e., a difference of 0.2). On the other hand, threshold-dependent metrics find it challenging to differentiate between neighboring noise intensities. Specifically, AUC-mROC struggles to discern algorithms with close noises. 

\begin{figure}[H]
  \centering
  \includegraphics[width=0.9\textwidth]{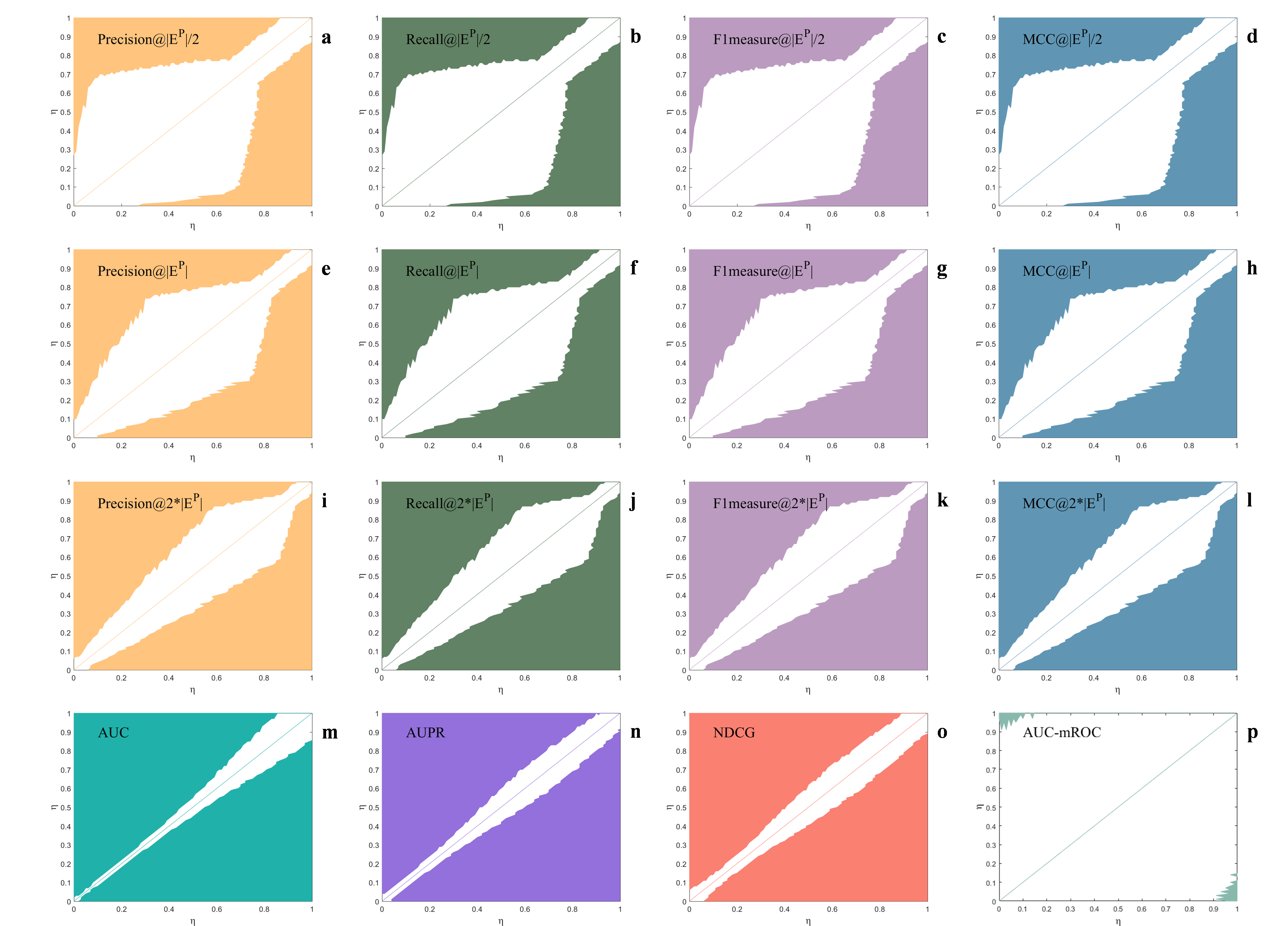}
  \caption{The binarized discrimination matrices of different evaluation metrics. The x-axis and y-axis represent the intensity of noise. (a)-(d), (e)-(h) and (i)-(l) respectively depict the results when the thresholds for Precision, Recall, F1-Measure, and MCC are \( |E^P|/2 \), \( |E^P| \), and \(2|E^P|\). (m)-(p) illustrate the outcomes for AUC, AUPR, NDCG, and AUC-mROC. }
  \label{heatmap}
\end{figure}

To provide a more intuitive comparison of the discrimination abilities of different metrics, we set \( p^{*}=0. 01 \). Only when \( p(\eta_1, \eta_2) < p^* \), we consider the current metric capable of distinguishing between noises \( \eta_1 \) and \( \eta_2 \). Figure \ref{heatmap} displays the binarized discrimination matrix. The colored areas represent elements for which \( p(\eta_1, \eta_2) < p^* \), while the white areas represent elements for which \( p(\eta_1, \eta_2) \geq p^* \). Clearly, a larger colored area (indicating the distinguishable region) suggests a stronger discrimination ability for the corresponding evaluation metric. Figure \ref{heatmap} shows that AUC, AUPR, and NDCG exhibit significantly superior discrimination abilities compared to other metrics. 

\section{Discussion}
We conduct a study on the discriminating abilities of evaluation metrics in link prediction on artificial networks. In a scenario where the likelihoods of links in the network are known, we devise a straightforward algorithm, whose accuracy can be regulated by adjusting a single parameter: the intensity of noise. By examining whether the evaluation metrics could accurately discern the performance differences of algorithms for different noise intensities, we can measure the discriminating ability of these metrics. Through observations on the magnitude of fluctuations in the metric values at given noise levels and the discrimination matrix composed of p-values, we found that the discriminating ability of AUC, AUPR, and NDCG was notably superior to other metrics. This includes commonly used BP (corresponding to Precision@k for \( k = |E^P| \)) and the recently proposed AUC-mROC. 

Link prediction in sparse networks is a classic imbalanced binary classification problem, where positive samples (i. e., missing links) are scarce while negative samples (i. e., non-existent links) are abundant. In such cases, to find as more as possible positive samples in limited attempts is more crucial than to put possible samples in relatively higher positions than negative samples. Both AUC-mROC and NDCG address this issue by assigning higher weights to the topper positions of the prediction list. However, the discriminating ability of AUC-mROC is found to be poor in the considered artificial networks studied in this paper. One potential reason could be the transformations applied to both coordinates of the ROC curve, making AUC-mROC curve highly sensitive to the accurate prediction of the initial entries in the prediction list. Given that \( q_{ij} \) in this study was independently generated from a uniform distribution, the average likelihood difference between positive and negative samples is relatively small, presumably much less than that in real-world networks. This implies that even with zero noise, the accuracy of the initial predictions made by the \( \Omega \) algorithm would not be exceptionally high. Therefore, if an evaluation metric is particularly sensitive to the accuracy of predictions at the very beginning of the list, its discriminating ability in this study would not be high. This also explains the poor discriminating abilities of other threshold-dependent algorithms when the threshold is relatively small. To further elucidate the discriminating abilities of weighted metrics, including AUC-mROC, it is essential to conduct more extensive research based on real networks and commonly used algorithms in the future. 

Finally, we strongly recommend readers to pay close attention to NDCG. While the discriminating ability of this metric is close to AUC and AUPR, it has been seldom utilized in previous studies of link prediction. As previously demonstrated, NDCG also assigns higher weights to the rankings at the forefront of the list in a logarithmic manner (albeit more conservatively than AUC-mROC), partially reflecting the demands of imbalanced classification. Therefore, we suggest readers to consider NDCG as an alternative evaluation metric. Furthermore, it is essential to highlight the limited discriminating ability of BP. Given its intuitive simplicity, BP still holds some reference value, however, relying solely on this metric to rank candidate algorithms might suppress the credibility of the conclusions drawn.

\end{document}